\documentstyle[11pt]{article}
\begin{document}
\thispagestyle{empty}
\noindent\hfill  OHSTPY-HEP-TH-93-014\\
\noindent\hfill\today \\
\begin{center}\begin{Large}\begin{bf} Light-Front QCD(1+1) Coupled to Adjoint
Scalar Matter\\
\end{bf}\end{Large}\vspace{.75cm}
 \vspace{0.5cm}
                       Stephen S. Pinsky \\[10pt]
                   {\em Department of Physics\\
                   The Ohio State University\\
                      174 West 18th Avenue\\
                       Columbus, OH~~43210}\\
\vspace{0.5cm}
                        Alex C. Kalloniatis \\[10pt]
                       {\em Max-Planck-Institut f\"ur Kernphysik \\
                         Postfach 10 39 80 \\
                         D-69029 Heidelberg \\}
\end{center}
\vspace{1cm}\baselineskip=35pt
\begin{abstract} \noindent We consider adjoint scalar matter
coupled to QCD(1+1) in light-cone quantization on a finite `interval' with
periodic  boundary conditions. We work with the gauge group SU(2) which is
modified to ${\rm{SU(2)/Z_2}}$ by the  non-trivial topology.
The model is interesting for various nonperturbative approaches
because it is the sector of zero transverse momentum
gluons of pure glue QCD(2+1), where the scalar field is
the remnant of the transverse gluon component. We use the
Hamiltonian formalism  in the gauge $\partial_- A^+ = 0$.
What survives is the dynamical zero mode of $A^+$, which in other theories
gives topological structure and degenerate vacua. With a point-splitting
regularization designed to preserve symmetry under
large gauge transformations, an extra $A^+$ dependent term appears in the
current $J^+$. This is reminiscent of an (unwanted) anomaly.
In particular, the gauge invariant charge and the similarly
regulated $P^+$ no longer commute with the Hamiltonian. We show
that nonetheless one can construct physical states of definite
momentum which are not {\it invariant} under large gauge
transformations but do {\it transform} in a well-defined way.
As well, in the physical subspace we recover vanishing
{\it expectation values} of the commutators
between the gauge invariant charge, momentum and Hamiltonian operators.
It is argued that in this theory the vacuum is nonetheless trivial and the
spectrum is consistent with the results of others who have treated
the large N, SU(N), version of this theory in the continuum limit.
\end{abstract}
\newpage\baselineskip=18pt

\def\bzm #1 {\langle  #1\rangle_\circ }                
\def\no {n\,} \def\ze {\circ\,} \def\sq {\star\,}      
\def\zml #1#2#3{\hbox{ \vbox{\ialign{##\crcr
    ${\,\scriptstyle #1\,}$\crcr
     \noalign{\kern1pt\nointerlineskip}
    $\displaystyle{#2}_{#3}$\crcr}}}}

\section{Introduction}  The unique features of `front form' or
light-cone quantized field theory provide a powerful tool for the study of QCD.
Of primary importance in this approach is the existence of a vacuum state
that is the ground state of the full theory. The existence of this
state gives a firm basis for the investigation of many of the complexities
that must exist in QCD. In this picture
the rich structure of vacuum is transferred to the zero modes of the
theory. Within this context the long range physical phenomena of
spontaneous symmetry breaking \cite{Rob93} as well as the
topological structure of the
theory \cite{KPP94} can be associated with the zero mode(s) of the fields in a
quantum field theory defined in a finite spatial volume and quantized at equal
light-cone time \cite{PaB85}.

These phenomena are realized in two quite different ways in several simpler
theories.  For example, spontaneous breaking of $Z_2$ symmetry  in
$\phi^4_{1+1} $ occurs via a {\it constrained} zero mode
\cite{Rob93}. There the zero mode satisfies a non-linear constraint
equation that relates it to the dynamical modes in the problem \cite{MaY76}.
At the
critical coupling a bifurcation of the solution occurs. These solutions in turn
lead to new operators in the Hamiltonian which break the $Z_2$
symmetry at and beyond the critical coupling.  The work of Franke et al.
\cite{FNP81} shows that such  constrained zero modes are present in gauge
theories, for example in (3+1) dimensions.   Quite separately, a dynamical zero
mode was shown in \cite{KPP94} to arise in pure $SU(2)/Z_2$ Yang-Mills in 1+1
dimensions.   A complete fixing of the gauge leaves the theory with one
degree of freedom, the zero or gauge mode of the vector potential $A^+$.
The theory has a discrete spectrum with zero momentum $P^+$ states
corresponding to modes of the flux loop around the finite space.
Only one state has an eigenvalue zero of
energy, $P^-$, and is the true ground state of the theory. The non-zero
eigenvalues are proportional to the length of the spatial box,
consistent with the flux loop picture.  This is a direct result of the
topology of the space. As the theory
considered there was a purely topological field theory the exact solution
was identical to that in the `instant form' approach on the analogous
spatial topology \cite{Het93}.

In the present work we consider $QCD_{1+1}$ coupled to scalar adjoint
matter, also studied  in the absence of zero modes by
\cite{DKB93}. This theory can be obtained by dimensional reduction to
(1+1) of pure glue theory in (2+1) dimensions. The scalar field is the
remnant of the transverse gluon component. Our study of this theory
is part of a long term program to attack QCD(3+1) through the zero modes
sectors starting with studies of
lower dimensional theories which are themselves zero mode sectors of higher
dimensional theories \cite{KPP94,KaP93}. A complete gauge fixing has recently
been found for QED \cite{KaP93} which further supports this program. In all
these cases, the goal was to disentangle the {\it de}pendent from the {\it
in}dependent degrees of freedom, in particular for the zero modes.
As we showed in an earlier treatment \cite{PKP95}, dimensionally reduced pure
glue theory in (2+1) dimensions exhibits both types of zero modes.
The dynamical zero mode comes from the (1+1) Yang-Mills sector
while the constrained mode is in the
scalar, namely remnant transverse gluon, field.  In \cite{Kall95}
a method for solving the, in this case, linear constraint was
developed with the result that there is no vacuum degeneracy
even though hints of how such degeneracy could take place appeared.
We shall comment more on this in the conclusions.
Here, we investigate the
consequences of regulating currents using gauge-invariant point-splitting
similar to that used by Manton \cite{man85} in the Schwinger Model and more
recently by Lenz, Shifman and Thies \cite{LST94} for
$QCD_{1+1}$ coupled to adjoint fermions. This regularization respects the
symmetries of the theory under large gauge transformations and Weyl
conjugations.
The results are somewhat surprising:  we find that an extra term is
generated in
the current $J^+$ whose diagonal color charge $Q_3$ itself is meant to
generate global gauge transformations. This `anomalous' term involves the
zero mode of $A^+$ left after a complete gauge fixing of the  theory.
A similar term appears in the momentum $P^+$ operator.
These contributions mean that symmetries such as Lorentz and
charge invariance cannot be realized in a Hilbert space of states
satisfying the large gauge symmetry.
This problem is in fact generic to this type of treatment of
any (1+1)-dimensional gauge theory on the light-cone.
It is peculiar because this does not appear to occur in quantization
on a space-like surface of the same theories. We propose that the
resolution to the dilemma is to give up {\it invariance} of the `physical'
states under large gauge transformations. The naive normal ordered
charge and momentum operators, which commute with the Hamiltonian,
can be used to label the states. Because large gauge transformations
can be realized in the quantum theory as unitary transformations of the
Hilbert space, the physics, such as the spectrum of eigenvalues of the
mass-squared operator, is invariant.

In section II we formulate the general structure of the theory in
$SU(2)/Z_2$. In
section III we quantize the model and introduce the zero mode structure of the
theory. We discuss the symmetries of the theory.  In Section IV we
perform the point-splitting regularisation for $Q_3$ and $P^+$.
In Section V we discuss our results and contrast it with the
treatments in \cite{PKP95,Kall95}.

\section{ Dimensional Reduction of Pure Glue Theory }
In the following we briefly reiterate the formulation we presented first in
\cite{PKP95}.  We take the pure Yang-Mills Lagrangian density
\begin{equation}
  {\cal L}= -{1\over2}{\rm Tr}\bigl({\bf F}^{\mu\nu}{\bf F}_{\mu\nu} \bigr)
   \ ,{\rm with} \quad
 {\bf F}^{\mu\nu} \equiv \partial^\mu{\bf A}^\nu -
\partial^\nu{\bf A}^\mu
   + ig \bigl[{\bf A}^\mu,{\bf A}^\nu \bigr]
\label{Lagr}
\end{equation}
for which the energy-momentum tensor is
$ \Theta^{\mu\nu} = 2{\rm Tr} \bigl({\bf F}^{\mu\kappa} {\bf
F}_\kappa^{\phantom{\kappa}\nu} \bigr)- g^{\mu\nu}{\cal L}$. In the front
form, it
is convenient to split the latter and their Lorentz indices
$\mu(\nu)$ into the longitudinal values $\alpha(\beta   )= +,-$ and into the
transversal components. We use the convention $x^\pm = (x^1 \pm x^2)/
\sqrt{2}$ and $A+=A^-$.The Lagrangian and the energy-density
$\Theta ^{+-}$ then disentangle nicely and in (2+1) dimensions we find,
 ${\cal L} = -{1\over2}{\rm Tr} \bigl({\bf F}^{\alpha
\beta}
 {\bf F}_{\alpha \beta} + 2{\bf F}^{\alpha j}{\bf F}_{\alpha j} \bigr)
  \quad {\rm{and}} \quad
  \Theta^{+-} =  {\rm Tr}\bigl({\bf F}^{-+}{\bf F} ^{-+}
\bigr)$ .
The original formulation of Discretized Light-Cone Quantization (DLCQ)
\cite{PaB85} was formulated in terms of  only the normal modes. Here
we pursue a complementary approach and analyze  the theory in terms of only the
zero modes. In particular we consider the theory with only transverse zero
modes by requiring
$\partial_i {\bf A}^\mu = 0$.   Since all fields are thus independent of the
transverse coordinate it is convenient to readjust units by scaling out the
transverse length
$L_{\!\bot}$. An adjustment of notation,
${\bf A}^\mu = ({\bf A} ^+,{\bf A}^-,{\bf A} ^1) \equiv ({\bf V},{\bf
A},\Phi)$,
helps to avoid too many indices. The model-theory then takes the form of a
(1+1)
gauge theory covariantly coupled to an adjoint scalar matter field
\cite{DKB93},
\begin{equation}
 {\cal L}={\rm Tr}\Bigl(-{1 \over 2}{\bf F}^{\alpha\beta}{\bf F}_{\alpha \beta}
    +{\bf D}^\alpha \Phi {\bf D}_\alpha \Phi \Bigr) \ .
\end{equation}
The equations of motions are correspondingly
\begin{equation}
 {\bf D}_\beta{\bf F}^{\beta \alpha} = g{\bf J }^\alpha
\ ,{\rm with} \quad
 {\bf J }^\alpha = - i\bigl[\Phi,{\bf D}^\alpha  \Phi
\bigr]
  \ ,\quad {\rm{and}} \quad
 {\bf D}^\alpha {\bf D}_\alpha  \Phi = 0 \ .
\label{eqofmot}
\end{equation}
The currents ${\bf J}^\alpha$ are introduced for convenience and are
only covariantly conserved.
To fix the gauge, we follow the procedure given in
\cite{KPP94} and find that
$A^+$ only has a zero mode, i.e. $   \partial_-{\bf A}^+ = 0$. After a
rotation in color space, the matrix is diagonal in color space.
In the instant form this gauge
has been used by \cite{Het93,LaS92} to name a few.  In a context
related to the front form it has also been used by \cite{LTL91}.

In the present notation
$  {\bf F}^{-+} = \partial_+{\bf V}-{\bf D}_-{\bf A} $. The first of our three
equations of motion in the gauge sector is thus simply Gauss' law,
${\bf D}_-{\bf F}^{-+} =-{\bf D}_-^2{\bf A} = g{\bf J }^+ $, realized here as a
second class constraint in the nomenclature of Dirac.
In the absence of gauge-fixing these are first class
constraints and are a consequence of the gauge-symmetry. With the gauge-fixing,
these can be realized as quantum operator constraints. The off-diagonal part of
this equation can be solved strongly yielding
\begin{equation}
  {\bf A} = -g{1\over {\bf D}_-^2}{\bf J }^+ \ .
\end{equation}
The diagonal projection
of the zero mode part of Gauss's law remains first class.
This must be satisfied as a condition on the states, {\it i.e.}
$\bzm{\bigl({\bf J }^+ \bigr)_{\rm diag}} \vert{\rm phys} \rangle = 0$.
Analogous constraints can be found in other contexts, eg \cite{MiP94}.
Here $\langle {f} \rangle_\circ$ is the zero mode projection $\int_{-L}^{L} dx
f(x)/2L$ of some quantity $f(x)$, as in our earlier work, eg \cite{PKP95}.
Since we pursue a Hamiltonian approach we do not
give the detailed expressions for the genuinely dynamical equations.
Sufficeth to say, they exist for the gauge mode ${\bf V}$ and for
the scalar field. The zero mode projection of the color-diagonal
part of the latter equation will be shown to
generate a constraint equation. In the Dirac procedure it occurs as a
second class constraint.

Finally, the formal expression for the Hamiltonian is
\begin{equation}
   P^- = \int _{- L}^{+ L} \!\! dx^- {\rm Tr}
   \   \bigl(\partial_+{\bf V}-{\bf D}_-{\bf A}\bigr) ^2
       = \int_{- L}^{+ L} \!\! dx^-{\rm Tr}
   \   \bigl(\ \partial_+{\bf V}  \partial_+{\bf V}
   - g^2{\bf J}^+{1\over{\bf D}_-^2}{\bf J }^+
   \  \bigr)  \ .
\label{hammat}
\end{equation}
It describes the interaction of two matter currents of adjoint scalars via an
instantaneous gluon-like interaction
\cite{DKB93}. The instantaneous gluon is modified by the zero mode of
${\bf A}^+ $. This zero mode structure produces an effective mass
\cite{pin94} and in no case is $ 1/{\bf D} _-^2 $ singular.

\section{ Quantization and Matter Currents } In the following we will use a
color helicity basis for all field matrices of the form
\begin{equation}
     \Phi = \tau^3 \varphi_3 + \tau^+ \varphi_+ + \tau^-
\varphi_-
\end{equation}
where $\tau^a =\sigma^a/2$ and $\tau^\pm =(\tau^1 \pm i
\tau^2)/\sqrt{2}$. The zero mode matrix ${\bf V} $ is diagonal, thus
 $ {\bf V} = v\ \tau_3$,
where $v \equiv v(x^+)$ is a quantum mechanical operator as discussed in
\cite{KPP94,man85}.   The conjugate momentum is
$  p  \equiv  \delta L / \delta v  = 2 L  \partial_+v
$. It satisfies the commutation relation
$   \bigl[v,p \bigr] =
   \bigl[v,\ 2 L \partial _+v \bigr]= i$.
Whenever we see the operator $ v $ in the subsequent analysis it is understood
that we work in a representation which diagonalizes the operator $ v $.
Thus $ p =
-i d/dv $, see \cite{man85}.

The diagonal components of the hermitean matrix $\Phi $ is denoted by
$\varphi_3$. Any real-valued boson field subject to
periodic boundary conditions can be represented by
\begin{equation}
   \varphi_3 (x^+,x^-) ={a_0(x^+) \over \sqrt{4\pi}}
   +{1 \over \sqrt{4\pi}} \sum _{n=1}^{\infty} w_n
    \Bigl(a_n(x^+){\rm e}^{-ik_n x^-}+
    a^\dagger_n (x^+){\rm e}^{+ik_n x^-} \Bigr)  \ .
\end{equation}
where $ k_n = 2\pi n /(2 L)$ denote the discretized momenta. With $
\bigl[a_n,a^\dagger_m \bigr]=
\delta_{n,m}$ ($n,m=1,\dots,\infty$)  and coefficients
$w_n=1/ \sqrt{n}$ one gets the correct commutation relations for
field operators with {\it no} zero modes,{\it i.e.}
\begin{equation}
    \Bigl[ \varphi_3(x) , \pi^3(y) \Bigr]_{x^+=y^+} =
    {i\over2} (\delta (x^- - y^-) - {1\over{2L}}) \ ,
\end{equation}
The momentum field conjugate to $\varphi_3$ is denoted by
$\pi^3 \equiv \partial_-\varphi_3$. The `zero mode' $a_0 = a_0^\dagger $,
however, obeys a constraint equation
obtained by the projection
${\rm Tr}\ \bzm{\tau^3{\bf D}^\alpha{\bf D}_\alpha \Phi } = 0$.
As we observed in our previous work \cite{PKP95}, this constraint is
{\it linear} in $a_0$, which appears through the currents $J^+_\pm$.
Thus this is quite different in structure from the constraint equation of
the $\phi^4_{1+1}$ theory  \cite{Rob93}. We return to this in the final
discussion.

The off-diagonal components of $\Phi$ are complex valued fields,
$\varphi_+ (x^+,x^-) = \varphi_-^\dagger (x^+,x^-)$. Any such boson field
subject
to periodic boundary conditions can be written as
\begin{equation}
   \varphi_- (x^+,x^-) ={1 \over \sqrt{4\pi} }
   \Bigl( \sum_{n=0}^{\infty} b_n u_n \ {\rm e}^{-ik_n x^-} +
          \sum_{n=1}^{\infty} d^\dagger_n\ v_n{\rm e}^{+ik_n x^-} \Bigr)\ .
\end{equation}
The analogy with the (complex) Dirac spin components is convenient for
discussing
some of the symmetries of the theory. In the following we take a
somewhat different approach to our treatment in \cite{PKP95}.
The canonical momentum fields conjugate to $\varphi_-$ and $\varphi_+$ are
$  \pi^- = \bigl(\partial_- + igv\bigr)\varphi_+ $ and
$  \pi^+ = \bigl(\partial_- - igv\bigr)\varphi_- $
and these satisfy equal $x^+$ commutation relations with the fields,
\begin{equation}
\Bigl[\varphi_-(x^-) , \pi^-(y^-) \Bigr] =
    \Bigl[\varphi_+(x^-) , \pi^+(y^-) \Bigr] =
    {i\over2} \delta (x^- - y^-)
\ .
\end{equation}
These relations can be satisfied with the choice of coefficients
\begin{equation}
    u_n\equiv{1\over\sqrt{\vert n + z \vert} }
   \ , \quad {\rm{and}} \quad
    v_n\equiv{1\over\sqrt{\vert n - z \vert} }
   \ , \quad {\rm{with}} \quad  z \equiv { gvL \over \pi}
\label{coeffu}
\end{equation}
and with commutation relations for the $b, d$ operators
\begin{eqnarray}
     \bigl[ b_n , b^\dagger_m \bigr] =
     {\rm{sgn}}(n+z) \delta_{n,m}, \quad
     \bigl[ d_n , d^\dagger_m \bigr] =
     {\rm{sgn}}(n-z) \delta_{n,m}, \quad
    \bigl[ b_n , d_m \bigr]  =  \bigl[ b_n , d^\dagger_m
\bigr] = 0
\ .
\label{focomrel}
\end{eqnarray}
Note that the assignment of creation and annihilation operator depends on $z$.
With these results it is useful to express the conjugate momentum
expansion as
\begin{equation}
  \pi^+(x^+,x ^-)={-i\over 4L}{\sqrt{4\pi}}
\Bigl(\sum_{n=0}^{\infty}
   b_n {{\rm{sgn}}(n+z) \over  u_n} {\rm e}^{-ik_n x^-} - \sum_{n=1}^{\infty}
  d^\dagger_n {{\rm{sgn}}(n-z) \over v_n} {\rm e}^{+ik_n x^-}\Bigr)\ .
\end{equation}
with $\pi^- = (\pi^+)^\dagger$. The result that
$(n+z) u_n = (n+z)/\sqrt{n+z} = {\rm {sgn}}(n+z)/u_n$ is useful for
obtaining this result. One should also observed that in the above we have
made a choice at the edge of the ``Dirac-sea" to assign to
the dynamical zero mode of
$\varphi_-$ a $ b_0$ rather than a
$d^\dagger_0$ operator.
One could write it as a superposition but a trivial Bogoliubov transformation
allows one to transform the vacuum and states between different choices.

Because of the torus geometry of our space and the non-Abelian structure of the
gauge group, there remain large gauge transformations which are still
symmetries of the theory \cite{Grib78}. We have explained how to
completely fix the gauge with respect to small gauge transformations.
The large gauge transformations are
generated by local $SU(2)/Z_2$ elements
$ V(x) = \exp({-i{{\pi} \over{ L }} x \; \tau_3)}$,
which is anti-periodic. On the diagonal component $v$ it generates shifts that
are best expressed in terms of the dimensionless $ z $:
$ z  \rightarrow  z ' =  z  + 1$.
A shift by any integer is generated by repeated application of this
transformation.   On the scalar adjoint
fields, the effect of the transformation is
\begin{eqnarray}
     \varphi_3 & \rightarrow & \varphi'_3 = \varphi_3 \\
     \varphi_\pm & \rightarrow & \varphi'_\pm =
\varphi_\pm
             \exp{ (\mp i{{\pi}\over{L}} x) }.
\end{eqnarray}
\noindent This leads to the following effect on the modes $b$ and $d$
\begin{eqnarray}
      d_{1}^\dagger & \rightarrow &  b_0 \\
      d_{n}^\dagger & \rightarrow & d_{n-1}^\dagger
\quad n \geq 2 \\
      b_n & \rightarrow & b_{n+1} \quad n \geq 0 .
\end{eqnarray}
As a consequence we find a spectral flow very similar to the problem with
fermions \cite{man85, LST94}: some of the hole states are elevated to occupied
particle states. However, the physical content of any given domain $ M
\leq z \leq M+1$ is the same for all
$M$. We shall label the domains by the integer $M$.

The theory has an additional Weyl conjugation symmetry,
$ z  \rightarrow  z ' = -  z$.
On the $b$ and $d$ modes the transformation  is similar to charge conjugation
\begin{equation}
b_n  \leftrightarrow  d_n \quad n
\geq 1 \ , \\ b_0  \leftrightarrow  b_0^\dagger .
\end{equation}
We see that the Weyl conjugation preserves the commutation relations in an
interesting way. The factor
${\rm{sgn}}(n+z)$ in the commutation relations changes sign for $n=0$ when $ z
\rightarrow - z $ and the interchange of $b_0
\leftrightarrow b_0^\dagger$ compensates for it. This symmetry also introduces
a
degeneracy in each domain, which we label by the integer $M$. The lower
half of the domains,
$ M \leq z \leq M+{1 \over 2} $  are related to the upper half
$ M + {1 \over 2} \leq z \leq M+1 $ of the domains. To see this consider the
$M=0$ domain, the fundamental modular domain (FMD).
The region $ 0 \leq z \leq{ 1\over 2}$ is equivalent to the
region  $-{1\over 2}
\leq z \leq 0$ by Weyl conjugation and this region is equivalent to the region
${1\over 2} \leq z \leq 1$ by a large gauge transformation.  This in effect
forces the domain to be symmetric about $z={1 \over 2}$. In
\cite{LST94} was shown that it is this symmetry about
${1 \over 2 }$ that gives $QCD_{1+1}$ coupled to adjoint fermions a degenerate
vacuum. In that model one vacuum wave function is centered  just above $z=0$
and
the other is centered just below $z=1$.

To close this section we repeat
the argument in \cite{PKP95} showing that the gauge mode
$ z $ can be written in terms of an explicitly color singlet object, namely the
Wilson loop constructed via a   contour ${\rm C}$ along the $ x $ direction
from $-
L $ to $ L $:
\begin{equation}
  W  ={\rm Tr}{\rm{P}} \exp(ig \int_{{\rm C}} dx_\mu
 {\bf A}^\mu ) ={\rm Tr} {\rm{P}} \exp (ig \int_{-L}^{+ L} d x {\bf A}^+ )
 ={\rm Tr} \exp(2\,i\, z  \,\pi \, \tau^3).
\end{equation}
Thus we can relate $ z $ to $ W $ modulo the integers,
$ z  ={1\over{2\pi}}{\rm{arcos}} ({{ W }\over2}) \;$.
The integer shifts are nothing but the Gribov copies discussed earlier. Observe
that the dynamical quantity $ W $ attains its minimum value at $ z = { 1
\over 2}$ making the point $z=\frac{1}{2}$ the symmetry point in
the FMD.

\section{Point-Splitting}
We begin by rewriting Gauss' law in components, {\it i.e.}
\begin{equation}
   -\partial_-^2 A_3 = g J^+_3, \qquad\quad
   -(\partial_- + ig v)^2 A_+
    = g J^+_+,
\label{gausscomp}
\end  {equation}
and the hermitian conjugate of the latter with
$(J^+_+)^\dagger \equiv J^+_-$. One would like to invert these to express
$A _3$ and $A _\pm$ in terms of the currents $J^+$ which, according to
Eq.(\ref{eqofmot}), are defined as
\begin{equation}
  J^+_3 ={1\over i} \bigl(\varphi_+ \pi_- - \varphi_-
\pi_+ \bigr)
  \  {\rm {and}} \quad
  J^+_+ ={1\over i} \bigl(\varphi_3 \pi_+ - \varphi_+
\pi_3 \bigr)  \ .
\end{equation}
The first of the Gauss equations (\ref{gausscomp}) can be solved only if the
zero mode $ \bzm{J^{+}_{3}} $ on the r.h.s vanishes also. This cannot be
satisfied as an operator, but rather as a condition on the
physical states, {\it i.e.}
$ \bzm{J^{+}_{3}} \vert{\rm{phys}} \ \rangle \ \equiv 0 $.
The calculation of the currents requires some care since it involves the
difference of the product of operators at the same point.  We regulate the
divergent sums by a gauge-invariant point-splitting.
\begin{eqnarray}
       J^+_3 =
                  \lim_{\varepsilon \rightarrow 0} \;
   {1\over i}\Bigl[
    \varphi_+(x^- - \varepsilon) \pi_-(x^-) {\rm e}^{-igv\varepsilon} -
    \varphi_-(x^- - \varepsilon) \pi_+(x^-) {\rm e}^{+igv\varepsilon}
    \Bigr]
\end{eqnarray}
This gives the follow form for the charge operator before performing
the sums or taking the limit
\begin{eqnarray}
 & Q_3 = - {\rm{sgn}}(z) b_0^\dagger b_0
\cos(\varepsilon \pi z/L) -
  {1 \over 2}{\rm e}^{i\varepsilon \pi z/L} \nonumber \\ +
\sum_{n=1}^{\infty}\Bigl(
  - & {\rm{sgn}}(n+ z ) b^\dagger_n b_n \cos(\varepsilon
\pi(n+ z)/L)
  - {1 \over 2}{\rm e}^{i\varepsilon \pi(n+ z)/L}+
\nonumber \\
    & {\rm{sgn}}(n-z) d^\dagger_ n d_n \cos(\varepsilon
\pi(n- z)/L)
  + {1 \over 2}{\rm e}^{i\varepsilon \pi(n-z)/L}
  \Bigr)
\ .
\end{eqnarray}
Performing the sums and taking the limit produces the following expression for
$Q_3$:
\begin{equation}
  Q_3 = -\sum_{n=0}^{\infty} {\rm{sgn}}(n+z) b^\dagger _n b_n
       + \sum_{n=1}^{\infty} {\rm{sgn}}(n-z) d^\dagger _n d_n
     +\;(z -{1\over 2})  \ .
\end{equation}
It is straightforward to show that the operator
$Q_3$ is symmetric under large gauge transformations and antisymmetric
under Weyl
conjugation.  We see the appearance of the anomalous term $z$ in $Q_3$.
It is easy to see that because of the kinetic term for the gauge
mode in the Hamiltonian Eq.(\ref{hammat}), the gauge invariant
regularized charge does not commute with the Hamiltonian.
It is important to point out that this does not occur
in a similar treatment of the charge operator in conventional
quantization, say, of the Schwinger model \cite{man85}.

We can also calculate $P^+$,
\begin{equation}
  P^+= \int_{-L}^{L} dx^- \bigl( \pi_3 \pi_3 + \pi_+
\pi_- +\pi_+ \pi_-\bigr)
\, .
\end{equation}
This operator involves operator products at the same point and requires the
same
careful treatment used to calculate $Q_3$. We find,
%
\begin{equation}
 P^+ = {\pi \over L} \Bigl( \sum_{n=1}^{\infty} \bigl(n \  a_n^\dagger a_n +
  |n-z| d_n^\dagger d_n + |n+z| b_n^\dagger b_n \bigr)
+ |z| b_0^\dagger b_0 -{1 \over 2}(z-{1 \over 2})^2 - {3 L^2 \over 2 \pi^2
  \varepsilon} \Bigr) \ .
\end{equation}
Direct calculation shows that $P^+$ is symmetric under large gauge
transformations and Weyl conjugation. The last term is a divergent constant
and can be trivially renormalized.
The term proportional to $(z - \frac{1}{2})^2$ however destroys the
commutativity of the momentum with the Hamiltonian.

One should also regulate noncommuting operator products in $P^-$
using point-splitting. This is too lengthy to treat here.
It suffices to say that the gauge factor introduced in the splitting could not
generate the terms required to maintain the commutation
relations between all the operators: only an extra term with
the conjugate momentum of the gauge mode can help
recover the vanishing commutators,
and this cannot arise from point-splitting.

We can however define an operator ${\tilde Q}_3$ via
\begin{equation}
Q_3 \equiv {\tilde Q}_3 + (z - {1\over2} )
\end{equation}
for which $[{\tilde Q}_3, P^-]$ = 0.
Moreover, we can similarly relate the regulated
momentum operator to a `naive' momentum operator ${\tilde P}^+$ via
\begin{equation}
P^+ \equiv {\tilde P}^+ + z {\tilde Q}_3 - {1\over2} (z - {1\over2})^2
\end{equation}
after subtracting the divergent constant.
One can then show that $[{\tilde P}^+,P^-] = 0$.

We have thus succeeded in constructing representations of the
charge and momentum operators in terms of a Fock space implementing
their symmetries with the Hamiltonian.
Evidently, the operators are not invariant under the large
gauge symmetries so that consequently the Hilbert space is not
invariant. The states can be labelled by
\begin{equation}
 | N_b; N_d; z\rangle = \Psi(z) | N_b;N_d \rangle
\end{equation}
reflecting the Fock-mode content as well as that of the gauge mode
in its ground state.
As mentioned in the introduction, only the ground state wavefunction
of the gauge mode contributes to the tensor product in the continuum
limit \cite{KPP94}.
Among all of these states we will restrict the
physical states to be those where
$ N_b $ = $ N_d $ in the FMD,
$0 < z < 1$ such that they are annihilated by the non-invariant
charge operator ${\tilde Q}_3$.
Under large gauge transformations these states transform.
For example, under $z \rightarrow z+1$, the charge ${\tilde Q}_3$
transforms ${\tilde Q}_3 \rightarrow {\tilde Q}_3 - 1$.
Thus if we represent the transformation by a unitary operator
$T$ and consider a physical state defined in the FMD
by $|N;N;z\rangle$, then
$ {\tilde Q}_3 T^\dagger |N;N;z\rangle =  T |N;N;z\rangle $.
Thus the transformed state is a state with an extra $d$-mode.
As the transformation can be represented by a unitary
operator on the Hilbert space constructed in the FMD
the spectrum of the mass-squared operator will be invariant.

In fact one can show that the expectation value of the
fully gauge invariant $Q_3$ vanishes between any Fock state
with equal numbers of $b$ and $d$ modes tensored with the
ground state of the gauge mode.
Since the theory is symmetric about $z=\frac{1}{2}$ the
wavefunctional will be either symmetric or antisymmetric.
We now take the expectation value
\begin{equation}
\int_0^1 dz |\Psi(z)|^2 \langle N;N| Q_3 |N';N'\rangle
=  \delta_{N,N'}  \int_0^1 dz |\Psi(z)|^2 (z-\frac{1}{2})
\end{equation}
which vanishes due to the product of the antisymmetric $(z-\frac{1}{2})$
with the symmetric $|\Psi(z)|^2$ under the integration.
Similarly, one can show that the commutators
$[P^-,Q_3]$ and $[P^-,P^+]$ vanish in the sense of expectation
values.

\section{ Discussion and Conclusions}
We considered the transverse zero mode sector
of $QCD_{2+1}$. The theory manifests itself as $QCD_{1+1}$ coupled to adjoint
scalar matter which has symmetries with respect to large gauge transformations
and Weyl conjugation. There were two zero modes upon complete gauge fixing.
One, the longitudinal zero mode of
$\varphi_{3}$, is constrained. The other, the zero mode of $A^+$, is a
dynamical field.  In the approach we took here, we generalized our previous
method of quantizing the theory. In \cite{PKP95,Kall95} the classical
Weyl and large gauge symmetries were not implemented in the quantum theory,
but rather extracted as classical phases in the field expansions employed
there.
This actually was rather convenient in that it permitted a
naive cutoff regularization which did not
violate this symmetry in the resulting quantum theory. Here we have realized
the
large gauge transformations in the quantum sense as well. The theory was
regularized using gauge-invariant point-splitting.
In this manner it became impossible to construct a Fock
representation of the charge and momentum operators respecting their
commutativity wih the Hamiltonian and, simultaneously, maintaining
the symmetry under large gauge transformations.
We found we could label states as eigenstates of the
parts of $Q_3$ and $P^+$ which transform under the
large gauge transformations. Correspondingly the states
themselves transform in this approach. In this way physical quantities, such
as the spectrum, remain invariant.

We now briefly discuss the significance of this work in relation to the
two previous papers \cite{PKP95} and \cite{Kall95}. In \cite{PKP95},
the potential governing the behaviour
of the gauge mode was computed explicitly. However, there remained a
logarithmic divergence. In \cite{Kall95}, it was shown that this
divergence could be removed by mass renormalisation.
The resulting potential is consistent
with the symmetry about $z=\frac{1}{2}$, argued in the present work by
looking at the Wilson loop. As mentioned,
the quantization in \cite{PKP95,Kall95} was performed keeping
the large gauge symmetry as a classical phase in the field expansions.
In this sense the spectral flow was not implemented in the quantum
Hilbert space.  It is not evident in the formulation of \cite{PKP95,Kall95}
how, in a theory with a chiral anomaly, that approach can
give the correct result,
given the picture of \cite{man85} relating the anomaly to spectral flow
in QED on a circle. Our findings here fill in that
gap while also recovering the result of  \cite{PKP95,Kall95}.
Work applying these methods to both
the Schwinger model and QCD(1+1) with adjoint fermions
is in progress \cite{PHR95}.

As for the spectrum of the theory, in the absence of the
analogue of $\theta$-vacua it would appear that the impact
of the gauge mode becomes minimal in the continuum
limit. The gauge mode wavefunctions are essentially
just sines or cosines, while in Fock-space matrix elements
the gauge mode only appears in denominators such as
$1/(n + z)$ or $1/(n-z)$. Consequently, in the limit of
large harmonic resolution $K$, essentially large integer momentum,
the impact of the gauge mode becomes small as has been observed
in \cite{PaB95}. In the absence of any richer structures in
this theory we are thus led back to the analogue of the original formulation
by Klebanov et al. \cite{DKB93}.
The possibility remains open that extension of this theory to
include more scalar fields and fermions, as would arise by a dimensional
reduction of QCD(3+1), would introduce enough richness so that the
the zero mode sector plays a more significant role in the
final spectrum of the theory.

\section{Acknowledgments} The authors acknowledge G. McCartor, Y. Frishman, B.
van de Sande , K. Harada and D.G. Robertson for advice.
(SSP) acknowledges the support of the U.S. Department of Energy.
(ACK) was supported by the DFG under contract DFG-Gz: Pa 450/1-2.
Travel support was provided in part by a NATO Collaborative Grant.

\end{document}